# Copycat: A High Precision Real Time NAND Simulator

Juyong Shin
Seoul National University
Department of Computer Science and Engineering
jyshin@archi.snu.ac.kr

Jongbo Bae
SK hynix
jongbo.bae@sk.com

Ansu Na
AIO co. Ltd.
asna@archi.snu.ac.kr

Sang Lyul Min
Seoul National University
Department of Computer Science and Engineering
symin@snu.ac.kr

## ABSTRACT
In this paper, we describe the design and implementation of a high precision real time NAND simulator called Copycat that runs on a commodity multi-core desktop environment. This NAND simulator facilitates the development of embedded flash memory management software such as the flash translation layer (FTL). The simulator also allows a comprehensive fault injection for testing the reliability of the FTL. Compared against a real FPGA implementation, the simulator's response time deviation is under 0.28% on average, with a maximum of 10.12%.

## 1. INTRODUCTION
NAND flash memory has several advantages over hard disk drives (HDDs) such as fast access time, low power consumption, resistance to vibration, and high level of parallelism. Advances in NAND flash memory technology has drastically increased capacity while lowering price per GB by shrinking cells, introducing the multi-level cell (MLC) technology that stores multiple bits in a single cell, and stacking cells vertically in a 3D structure [1]. With these advances, storage devices based on NAND flash memory are replacing HDDs not only in mobile devices where low power consumption and small form factor are important but also in server systems where low latency and high bandwidth are required. With the rapid expansion of flash memory based storage devices, it is critical to provide a development environment that achieves a faster time-to-market.

NAND flash memory does not allow in-place update and is subject to various faults [2]. In order to overcome these limitations, flash based storage systems use a software module called the flash translation layer (FTL). In many cases, the FTL and the underlying hardware of a flash storage system are developed in tandem as they are dependent on each other. This requires programming, debugging, and testing on a specialized hardware board that are more difficult when compared against software development for application programs.

This paper describes a NAND simulator called Copycat that meets the following requirements:

1. Easy accessibility: the simulator should be easily accessible to FTL developers; preferably it should run on a commodity desktop environment.
2. High precision: the simulator should model the NAND system's timing as precisely as possible.
3. Real time: the simulator should provide a real time response to allow for an accurate in vivo performance assessment of the FTL.
4. Fault injection capability: the simulator should be able to inject various faults in flash memory in a configurable manner to test the reliability of the FTL.

The rest of the paper is organized as follows: Section 2 gives background information and related work; Section 3 describes the design and implementation of the Copycat simulator; Section 4 presents results from a validation experiment using a real hardware; and finally Section 5 concludes and gives future research directions.

## 2. BACKGROUND AND RELATED WORK
### 2.1 NAND Flash Memory
A NAND flash memory chip consists of a set of blocks, each of which in turn consists of a set of pages. It supports three basic operations: read, program, and erase. The read and program operations return the contents of the page and write the supplied data to the page, respectively. The architecture of NAND flash memory does not allow in-place update of data and all the pages in a block must be erased before they can be programmed. Figure 1 shows the steps involved in the three NAND flash operations.

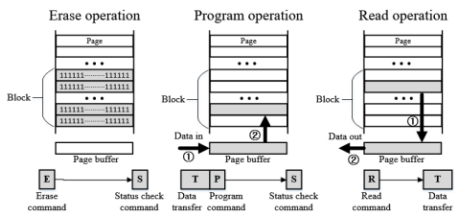

**Figure 1. Basic flash operations.**

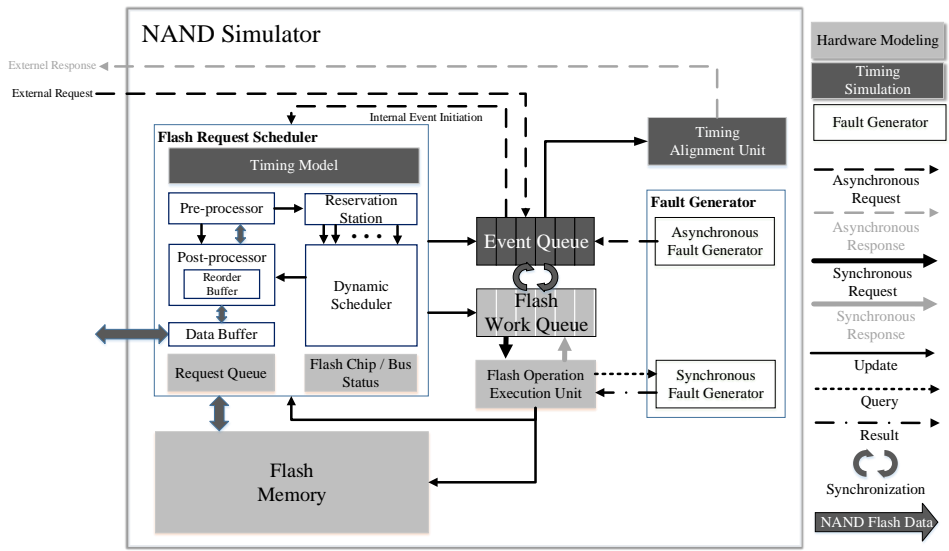

**Figure 2. Overall architecture of the Copycat NAND simulator.**

- **The erase block operation** consists of two phases. First, an erase command is issued along with the address of the block to be erased. After the erase is complete, a status check command is issued to detect any errors that might have occurred during the operation.

- **The program page operation** consists of two phases. First, the target page address is selected and the data to be written is transferred over the flash memory bus to the internal page buffer in the NAND flash memory chip. Then, a program command is to program the data. As in the erase block operation, a status check command is issued when the program is complete.

- **The read page operation** consists of two phases. First, a read command is issued along with the address of the page. This loads the page into the memory's internal page buffer and then the data in the page buffer is read out over the flash memory bus.

NAND flash memory cannot be used as a storage device in a straightforward manner because it does not allow in-place update of data. Thus storage devices based on NAND flash memory such as solid state drives (SSDs) and USB flash drives use a software layer called the flash translation layer (FTL) to mask this limitation.

A simple FTL that generates flash memory requests one at a time requires only a very primitive flash controller. However, a more practical FTL generates multiple concurrent flash memory requests and requires a flash controller that can service them in parallel by exploiting multi-chip parallelism. One such flash controller is Ozone [3] that executes flash operations in an out-of-order manner similar to the execution of instructions in modern superscalar microprocessors.

## 2.2 NAND Simulator

One of the early attempts to simulate NAND flash memory is reported in [4] where the authors modified an HDD simulator called DiskSim to simulate flash SSDs. They used the modified simulator to explore various options in designing SSDs. FlashSim is another SSD simulator based on the object-oriented programming paradigm. It was used to study performance and energy consumption variation of different FTLs in the SSD [5].

In both simulators, coarse-grained timing models were used for NAND flash memory, which was refined in NANDFlashSim [6] where a configurable timing model based on the microarchitecture of NAND flash memory was used. Microarchitectural features considered in NANDFlashSim include latency variation due to MLC, multi-die, multi-plane, and cached operations. One recent attempt to provide a real time feature in SSD simulation is VSSIM [7]. It is based on the QEMU/KVM virtual machine and allows for on-the-fly simulation of SSDs to study the effects of various SSD design parameters on the host performance.

## 3. COPYCAT NAND SIMULATOR

In this section, we describe the design and implementation of the Copycat simulator that supports a multi-bus/multi-chip flash system along with a flash controller based on Ozone [3]. In describing Copycat, we place an emphasis on how we fulfill our four requirements explained in the introduction.

Figure 2 gives an overview of the Copycat simulator. The simulator is based on event-driven simulation augmented by a real-time feature. Events in the Copycat simulator include:

1. Arrival of new flash memory requests from the FTL.

2. Departure of completed flash memory requests (i.e., acknowledgements) to the FTL.

3. Request movements between different queues in the Ozone out-of-order flash controller.

4. Initiation and completion of different phases of flash operations (read, program, erase).

Similar to other event-driven simulators, Copycat maintains a queue of events (**Event Queue** in Figure 2) sorted by the simulated time. The simulator repeatedly removes and processes the first event in the event queue. The processing time of the event is obtained from the **Timing Model** in the figure. The processing of an event may generate other future events, which are inserted into the event queue for later processing.

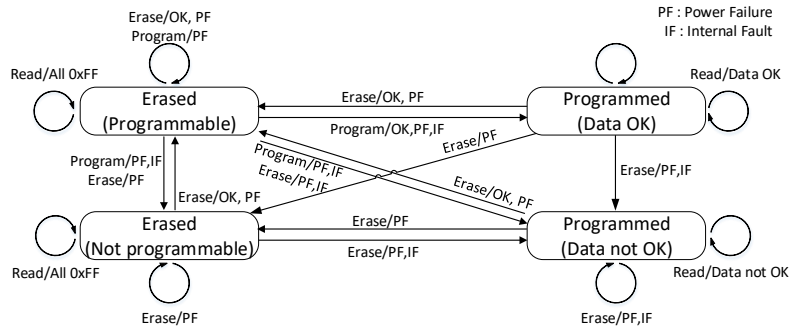

**Figure 3. Abstract fault model for NAND flash memory**

One key difference between the traditional event-driven simulator and Copycat is that each processing of an *externally visible event* (i.e., acknowledgement to the FTL after a flash memory request is completed) is accompanied by synchronization with real time by the **Timing Alignment Unit** shown in Figure 2. In this way, the external observer, i.e., the FTL, cannot distinguish the Copycat simulator from a real flash hardware. This synchronization with real time will be explained in Section 3.1.

Another difference is the exception that flash operations whose processing involves bulk data reads/writes from/to main memory are handled separately according to the earliest deadline first (EDF) scheduling. We will explain why this exception is needed and the associated EDF scheduling in Section 3.2.

The final difference is that the Copycat simulator is equipped with fault injection capability that models various faults that occur in flash memory. This fault injection feature is for testing the reliability of the FTL when subject to faults and will be explained in detail in Section 3.3.

## 3.1 Synchronization with Real Time

"Real time" and "high precision" are two key requirements of the Copycat simulator. These two requirements are met in Copycat by the timing alignment unit that is implemented by a separate thread on a dedicated core. The thread uses a CPU cycle counter, one of performance counters commonly available in modern microprocessors, for delaying an externally visible event with nanosecond accuracy. It repeatedly reads the CPU cycle counter and compares it against the completion time of the event (converted to CPU cycles) until the target time is reached.

## 3.2 EDF Scheduling of Flash Operations

In Copycat, the main simulation loop proceeds as follows:

1. Remove the first event in the event queue
2. Process the current event
3. Add to the event queue future events that are triggered by the current event
4. If the current event is an externally visible one, pass it to the timing alignment unit in an asynchronous manner and go to Step 1. In the background, the timing alignment unit delays the release of the event to the FTL until the real time reaches the simulated completion time.

One exception to this processing flow is for flash read and program operations that involve bulk data reads/writes from/to main memory that emulates the flash memory. For them, the event-driven scheduler inserts the needed data transfer tasks into the **Flash Work Queue** shown in Figure 2. The actual data transfers arising from simulating the flash operations are performed by the **Flash Operation Execution Unit**. The flash operation execution unit is again implemented by a separate thread running on a dedicated core to provide deterministic timing guarantee on the completion time of data transfers.

The flash operation execution unit performs data transfers in the earliest deadline first (EDF) order where the deadline is equal to the simulated completion time. After the data transfer is completed, the flash operation execution unit inserts the corresponding event to the event queue of the main event-driven simulator.

One complication in the EDF scheduling of data transfers is that the (simulated) completion time of a data transfer can be dynamically changed after it is submitted to the flash work queue. Such a situation can occur when there is a new external flash request that involves a data transfer whose completion time is earlier than those of some of previously scheduled ones. In this case, the completion times of data transfers in the flash work queue are updated accordingly but we do not preempt the ongoing data transfer even if its completion time is later than that of the new request. Note that such a handling does not violate the real time property as long as both data transfers are completed before their completion times since they will be correctly ordered when they are delayed in the timing alignment unit.

## 3.3 Fault Injection Capability

NAND flash memory is subject to various types of fault. There are two types of fault: internal and external ones. Internal faults occur since flash blocks fail over time, which is indicated by an error code returned by the status check command at the last phase of an erase or a program operation. An erase operation reports an error when one or more bits are stuck at 0, and thus cannot be reset to 1. Similarly a program operation reports an error when the number of bit difference between the programmed target page and the page buffer exceeds a threshold. These two faults are permanent and the blocks involved should be mapped out by the FTL and replaced by reserved spare blocks. Unlike internal faults, an external fault is not associated with flash operations. Power failure is a typical example of an external fault and it can occur at any time.

Internal and external faults can lead to anomalous behaviors in NAND flash memory. For example, power failure during a program operation can leave the target page in an indeterminate state. In the Copycat simulator, we use the abstract fault model in [8]. The abstract fault model associates a non-deterministic finite state machine with each page. Figure 3 shows an example of such a non-deterministic finite state machine used in the abstract fault model. This finite state machine makes the weakest assumption on the possible outcomes when a flash operation is subject to an internal fault (indicated by IF in figure) or a power failure (indicated by PF in figure). For example, when there is a power failure during a program operation, the target page can be in any of the following four possible states:

1. Erased (Programmable): This page contains all 1's and can be programmed.
2. Erased (Not programmable): This page cannot be programmed although it contains all 1's.
3. Programmed (Data OK): This page contains valid data.
4. Programmed (Data not OK): This page is corrupted and contains invalid data.

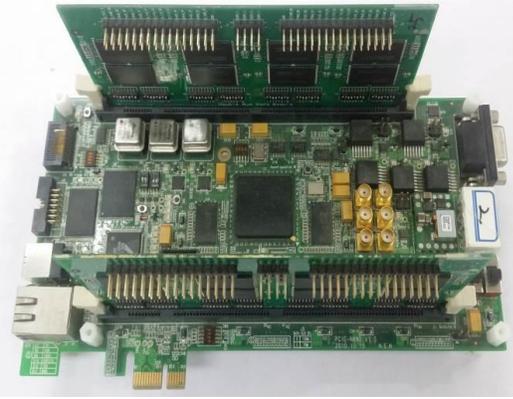

**Figure 4. In-house development board.**

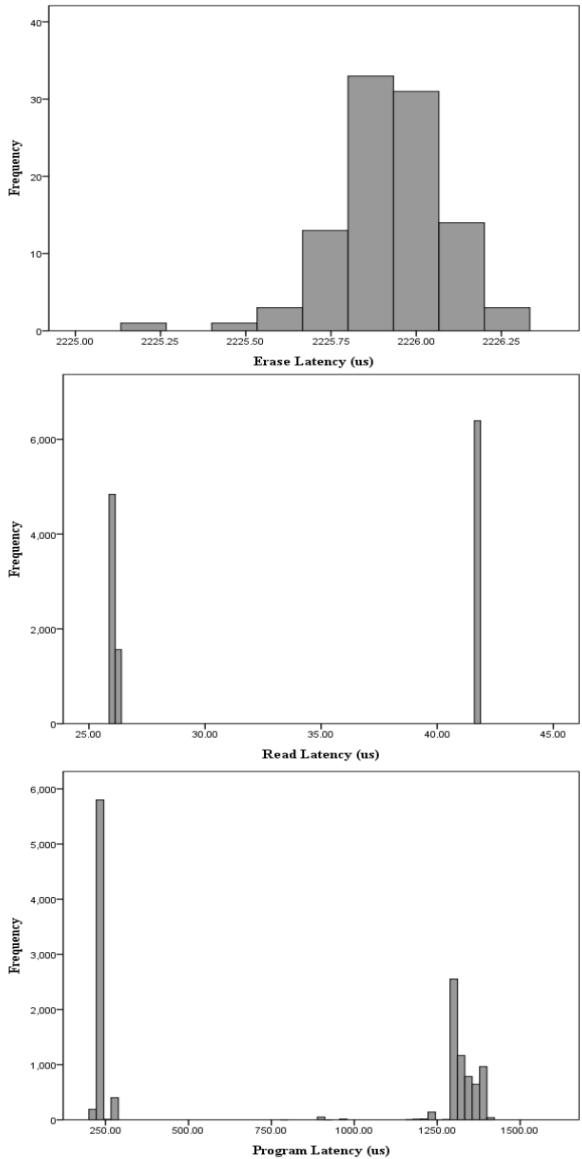

**Figure 5. Latency distributions of erase/read/program operations.**

In the Copycat simulator, internal faults are injected by the flash operation execution unit in a probabilistic manner whose specification is given by a configuration file. On the other hand, external faults (i.e., power failures) are injected by inserting the corresponding event into the event queue. If a fault occurs for an erase or a program operation, the abstract state associated with the target page is updated according to the transitions in the non-deterministic finite state machine used. This abstract state is used to service later reads from the same page.

## 4. EXPERIMENTAL RESULTS

The Copycat NAND simulator was implemented on a machine with an Intel multi-core microprocessor (Xeon E5-1620v2) running Linux 2.6.38. The machine runs at 3.7GHz and has 1MB L2 cache, 10MB L3 cache, and 128GB DRAM running at 1866MHz.

All the threads in Copycat including those running on dedicated cores were implemented as kernel threads in Linux. The CPU cycle counter we used was the time stamp counter (TSC) available on all modern x86 microprocessors.

For validation purposes, we also implemented an out-of-order NAND flash controller similar to Ozone [3] using an in-house development board shown in Figure 4. The development board has a Spartan6 FPGA (XC6SLX150T) and two NAND slots, each supporting two flash memory buses. We used a Toshiba 32Gbit MLC NAND flash chip (TH58NVG5D1DTG20) consisting of 4096 blocks, each with 128 pages of size 4Kbytes. The flash memory bus is 8 bits wide and operates at 33MHz. The latency time distributions of read/program/erase operations of the NAND flash chip are shown in Figure 5. Note that the latency times of read and program operations have bimodal distribution since in an MLC NAND flash memory reading the second bit in a cell takes a longer time than reading the first bit.

Figure 6 plots the response times of NAND flash memory requests from Copycat and those from the hardware implementation for a sample trace and Figure 7, the timing errors between the two. The response time deviation is under 0.28% on average, with a maximum of 10.12%. In Figure 7, there are a few spikes that cause the maximum deviation. A careful inspection of the completion times of events in the two systems (Copycat and the hardware implementation) revealed that those spikes were caused by a different ordering of data transfers for two or more flash operations that were requested at about the same time. Such situation cannot be avoided considering variable latencies even for the same flash operation and does not affect the response times of later flash memory requests in any significant manner.

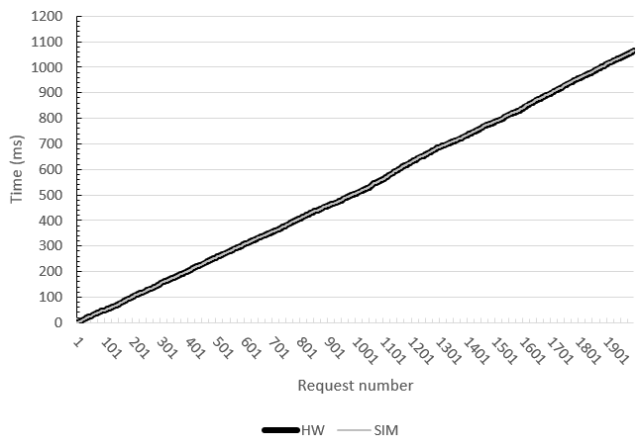

**Figure 6. Response time comparison between Copycat and hardware implementation.**

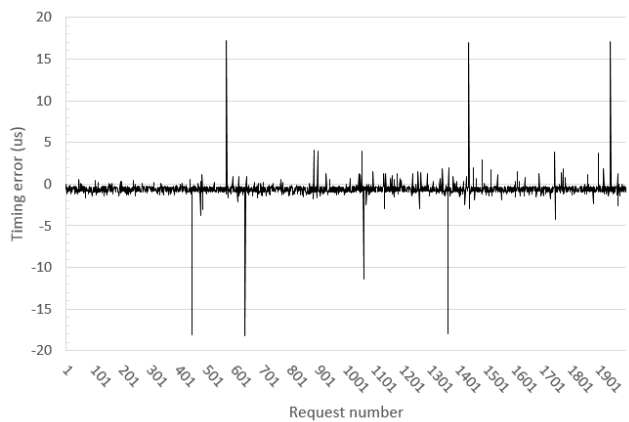

**Figure 7. Timing error between Copycat and hardware implementation.**

## 5. CONCLUSION

In this paper, we have presented a NAND simulator called Copycat. Besides being high precision and real time, the Copycat NAND simulator runs on a commodity multi-core desktop environment to facilitate easy programming, debugging, and testing of the flash translation layer (FTL). Moreover, Copycat provides a configurable fault injection capability for testing the reliability of the FTL.

Validation using an FPGA implementation NAND flash controller combined with real NAND flash memory chips showed that the Copycat simulator's response time deviation is under 0.28% on average, with a maximum of 10.12%.

We plan to extend the real time simulation technique in this paper to the whole solid state drive (SSD) system. Another interesting future research direction is to apply the same real time simulation to other I/O devices such as wired and wireless communication systems.